\documentclass[a4paper,12pt]{article}
\usepackage{graphicx}
\begin{document}

\title{Deep laser cooling in optical trap: two-level quantum model}

\author{O.N.Prudnikov$^{1,2}$ \thanks{E-mail: oleg.nsu@gmail.com}, A.V. Taichenachev$^{1,2}$, V.I. Yudin$^{1,2}$ and E.M. Rasel$^3$\\
{\em 1. Institute of Laser Physics, 630090, Novosibirsk, Russia} \\
{\em 2. Novosibirsk State University, 630090, Novosibirsk, Russia} \\
{\em 3. Institut f\"{u}r Quantenoptik, Universit\"{a}t Hannover}, \\
{\em Welfengarten 1, D-30167 Hannover, Germany}} \maketitle

\begin{abstract}
We study laser cooling of $^{24}$Mg atoms in dipole optical trap with pumping field resonant to narrow $(3s3s)\,^1S_0 \rightarrow
\, (3s3p)\,^{3}P_1$ ($\lambda = 457$ nm) optical transition. For description of laser cooling of atoms in the optical trap with
taking into account quantum recoil effects we consider two quantum
models. The first one is based on direct numerical solution of
quantum kinetic equation for atom density matrix and the second
one is simplified model based on decomposition of atom density
matrix over vibration states in the dipole trap. We search pumping
field intensity and detuning for minimum cooling energy and fast
laser cooling.
\end{abstract}

Pacs { 32.80.Pj, 42.50.Vk, 37.10.Jk,37.10.De}

\section{Introduction}
Nowadays deep laser cooling of neutral atoms is routinely used for
broad range of modern quantum physics researches including
metrology, atom optics, and quantum degeneracy studies. The
well-known techniques for laser cooling below the Doppler limit,
like sub-Doppler polarization gradient cooling \cite{Dalibard1989},
velocity selective coherent population trapping
\cite{Aspect1994,Adams1995} or Raman cooling
\cite{Kasevich1992,ctannoudji1995} are restricted to atoms with
degenerated over angular momentum energy levels or hyperfine
structure. However, for atoms with single ground state $^{24}$Mg,
$^{40}$Ca, $^{88}$Sr, $^{174}$Yb are of interest for developing
optical time standard these techniques can not be applied directly.
For example, for $^{24}$Mg atoms with the ground state $^1S_0$ the
Doppler cooling temperature ($k_B T_D \approx \hbar \gamma/2$) can
be reached on closed singlet transition $^1S_0 \rightarrow \,
^{1}P_1$ ($\lambda = 285.3$ nm). For lower temperature additional
cooling on $^3P_2 \rightarrow \, ^{3}D_3$ optical transition with
degenerated over angular momentum energy levels can be applied
\cite{Rasel2015,Rasel2012}. However, the experimental realization of
laser cooling on $^3P_2 \rightarrow \, ^{3}D_3$ optical transition
does not result significant progress. The atoms were cooled to
temperature  $T \approx 1\, mK$ is about Doppler limit only
\cite{Rasel2012}. The quantum simulation of laser cooling are also
shows the limitation of cooling temperature to about Doppler limit in
conventional MOT, formed by laser waves with circular polarization
\cite{Prudnikov2016}.

An alternative way of deep laser cooling of these elements is to use
narrow lines and ``quenching'' techniques of narrow-line laser
cooling \cite{holberg2001,Rasel2001,Rasel2003} successfully applied
for $^{40}$Ca atoms but, to our knowledge, still do not show
significant progress for $^{24}$Mg atoms.

Recently, laser-driven Sisyphus-cooling scheme was proposed for
cooling atoms in optical dipole trap \cite{Ivanov2011}. This scheme
utilize the difference in trap-induced ac Stark shift for ground and
exited levels of atom coupled by resonant laser light. The laser
cooling scheme has clear semiclassical interpretation: been excited
by resonant laser light on the bottom of shallow optical potential
related to the ground state an atom moves further in steepest
potential related to excited state. Spontaneous emission returns it
back to the shallow potential in the ground state. The loosing a
portion of energy in each act of this process results to atom
cooling after several cycles due to ``Sisyphus effect''
\cite{Ivanov2011}. This semiclassical model was applied for
description of laser cooling of Yb and Sr in optical dipole trap.

\begin{figure}[t]
\begin{center}
\includegraphics[width= 3 in]{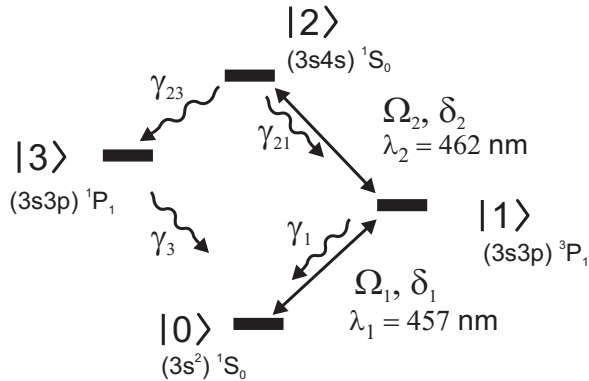}
\end{center}
\caption{\em Relevant energy levels for optical quenching and
cooling of $^{24}$Mg.} \label{mg_levels}
\end{figure}

In the following paper we study application of this cooling scheme
to $^{24}$Mg atom on narrow $(3s3s)\,^1S_0 \rightarrow \,
(3s3p)\,^{3}P_1$ ($\lambda_1 = 457$ nm, $\gamma_1 = 196$ s$^{-1}$)
optical transition. Additional light field resonant to
$(3s3p)\,^3P_1 \rightarrow \, (3s4s)\,^{1}S_0$ optical transition
($\lambda_2 = 462$ nm, $\gamma_{21} = 109$ s$^{-1}$ and $\gamma_{23}
= 2.1\cdot 10^7$ s$^{-1}$) is applied for optical quenching (see
Fig.\ref{mg_levels}), i.e. increasing the effective linewidth of
optical transition \cite{Rasel2003}. We find the semiclassical
description of laser cooling of Mg atom with narrow optical
transition can't be used here. For description of laser cooling we
use quantum approaches that allow to take into account optical
pumping and photon recoil effects in laser cooling process. In the
paper we point our attention to minimum laser cooling temperature
for described scheme and cooling time as well.

\begin{figure}[t]
\begin{center}
\includegraphics[width= 4.8 in]{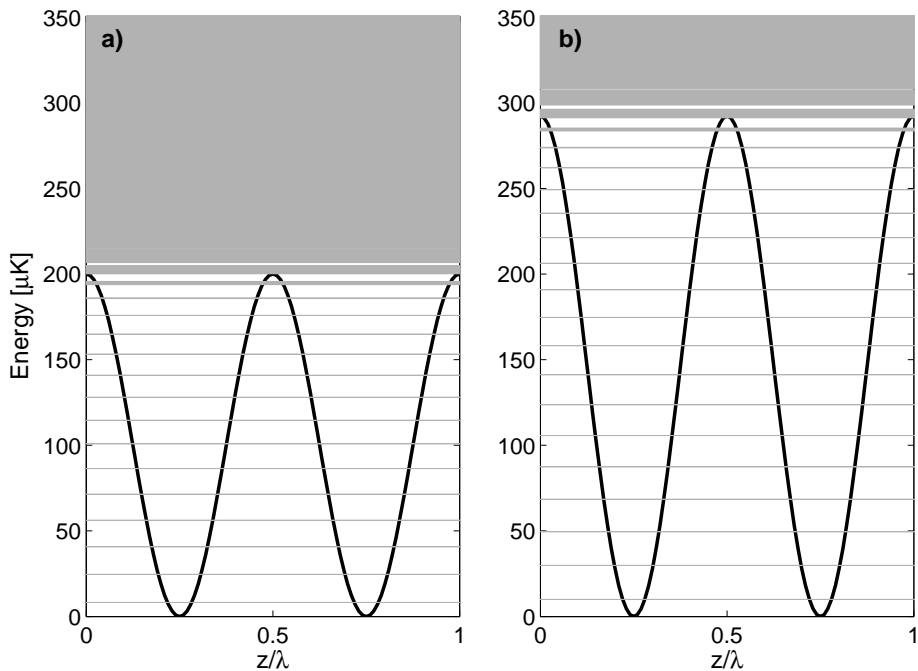}
\end{center}
\caption{\em Ac Shtark shift and vibration energy levels for ground
$^1S_0$ (a) and excited $^3P_1$ (b) states of $^{24}$Mg atoms in
dipole optical trap with $\lambda_D = 1064$nm.} \label{egen_states}
\end{figure}

\begin{figure}[t]
\begin{center}
\includegraphics[width= 3 in]{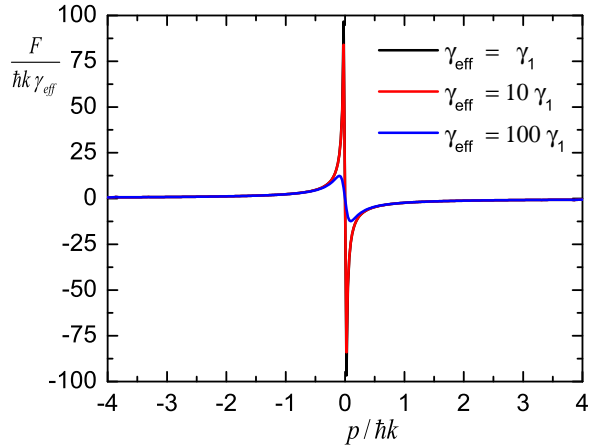}
\end{center}
\caption{\em Force on moving $^{24}$Mg atoms in the dipole trap
($U_g = 200\, \mu K$, $U_e = 292\, \mu K$) for different
$\gamma_{eff}$. Here $\delta_1 = 0$, Rabi of pumping field
$\Omega_1/\gamma_1 = 20000$ ($I \simeq 34$ $W/cm^2$).}
\label{mg_force}
\end{figure}

\section{Description of the model}
We consider the motion of $^{24}$Mg atom in the dipole optical trap
with $\lambda_{D} = 1064 \,nm$ that provide higher polarizability of
atom in the excited state $(3s3p)\,^{3}P_1$ than in the ground state
$(3s3s)\,^1S_0$. In the following paper we restrict our
consideration by two-level model assuming the quench field results
to increasing effective linewidth of optical transition to
$\gamma_{eff}$ \cite{Rasel2003}:
\begin{equation}
\gamma_{eff} = \gamma_1 + \gamma_2 \frac{\Omega_2^2}{\gamma_2^2+4\,
\delta_2^2}\, ,
\end{equation}
where  $\Omega_2$ is Rabi and  $\delta_2$ is detuning of quench
field. Thus, for example, to get $\gamma_{eff} = 100 \gamma_1$ at
$\delta_2 = 0$ one have to apply the quench field intensity $I_{q}
\approx 1.6 \,W/cm^2$. The simulated polarizability difference for
the optical dipole trap wavelength is about $\alpha_e/\alpha_g =
1.46$. In the trap (fig.\ref{egen_states}) the quantum nature of
atomic motion becomes essential. For considered optical trap depth
of the ground state $U_g = 200\, \mu K$ with vibration energy
separation of the lowest states are $\hbar \omega_g \approx 16.6\,
\mu K$. The excited state optical potential depth $U_e = 292\, \mu
K$ and the lowest states separation are $\hbar \omega_e \approx 20.1
\,\mu K$. The large energy levels separation require quantum model
for description of laser cooling dynamics of atoms in the trap.
Really, the semiclassical models can't be applied here because of
the velocity range of the semiclassical damping force on Mg atoms
has sharp variation in momentum space $\Delta p \ll \hbar k$  (see
figure \ref{mg_force}). As well the semiclassical parameter
$\varepsilon_R = \hbar k^2 /(2M\gamma) \gg 1$ is not small, that
also contradicts requirements for semiclassical approach
\cite{Dalibard1989,Juha1990,Prudnikov2004}.

For description of laser cooling of Mg in the optical trap we
consider two quantum approaches. The first one is based on
decomposition of atom density matrix on optical potential vibration
level states. Restricting by limited number of lowest vibration
states we simulate the stationary distribution over the vibration
levels in the trap, as well as the laser cooling dynamics to steady
state distribution. This approach is similar to method was described
in \cite{Castin1991}. However, in our model we also take into
account the optical coherence of different vibration states.

The second method we consider is based on direct numerical solution
of quantum equation for atom density matrix that allows to take into
account not only the fixed number of the lowest vibration level
states but whole density matrix of atoms, which also include
tunneling effects and above barrier motion. However, in this method,
due to the high complicity of the problem we omit the recoil effects
from the pumping field that is equivalent to orthogonal orientation
of wave vectors of pumping and optical trap light waves in one
dimensional model.

\subsection{Two-level model: exact numerical solution of quantum density matrix equation}
We consider the motion of Mg atom in the optical dipole trap is
standing light wave propagating along $z$ direction with linear
polarization along $x$. The pumping light field also linear
polarized along $x$ with wavevector along $z$ or $y$. The quantum
equation for atom density matrix describes evolution of internal and
external states of atoms
\begin{equation}\label{QQE}
\frac{\partial}{\partial t}{\hat \rho} = -\frac{i}{\hbar}
\left[{\hat H}_0 + {\hat V}_{ed},{\hat \rho}\right] +{\hat
\Gamma}\left\{{\hat \rho}\right\}
\end{equation}
with ${\hat H}_0$ is Hamiltonian, ${\hat V}_{ed}$ describes
interaction with pumping field and ${\hat \Gamma}\left\{{\hat
\rho}\right\}$ describes relaxation of density matrix due to
spontaneous decay.

 As was mentioned above, we restrict our consideration by effective two-level model with $(3s3s)\,^1S_0 $ is the ground (g) and
$(3s3p)\,^{3}P_1$ is excited state (e), assuming the influence of
the quench field $\Omega_2$ results to adjustable linewidth by
modification of decay rate from $\gamma_1$ to $\gamma_{eff}$ only,
as described in \cite{Rasel2003}. Further in the paper we omit
parameters indexes $\Omega_1$, $\delta_1$ and $\gamma_1$ by writing
$\Omega$, $\delta$ and $\gamma$ instead. The Hamiltonian of atom in
the trap has the form:
\begin{equation}
{\hat H}_0 = \frac{{\hat p}^2}{2M} +\hbar \omega_g(z) |g\rangle
\langle g | +\hbar [\omega_e(z) +\omega_0]|e\rangle \langle e |
\end{equation}
with optical potentials in the ground $\hbar \omega_g(z) = U_g
\cos^2(kz)$ and $\hbar \omega_e(z) = U_e \cos^2(kz)$ in excited
states ($\hbar \omega_0 = E_e-E_g$ is energy difference of
unperturbed ground and excited states). The wavevector $k =
2\pi/\lambda_{D}$ is defined by the dipole trap. Applying rotating
wave approximation the equation for atom density matrix components
in coordinate representation ${\hat \rho}(z_1,z_2)$ takes the
followign form:
\begin{eqnarray}\label{basic_eq}
\left(\frac{\partial}{\partial t} -\frac{i\hbar}{M}
\frac{\partial}{\partial q} \frac{\partial}{\partial z}\right)
\rho^{ee} &=& -\gamma_{eff}\, \rho^{ee} +i\frac{U_e}{\hbar}
\sin(2kz)\sin(kq) \rho^{ee}-\frac{i}{\hbar} \left[ {\hat V}
\rho^{ge}
- \rho^{eg} {\hat V}^{\dagger} \right] \nonumber \\
\left(\frac{\partial}{\partial t} -\frac{i\hbar}{M}
\frac{\partial}{\partial q} \frac{\partial}{\partial z}\right)
\rho^{gg} &=& {\hat \gamma}\{ \rho^{ee} \} +i\frac{U_g}{\hbar}
\sin(2kz)\sin(kq) \rho^{gg}-\frac{i}{\hbar} \left[ {\hat
V}^{\dagger} \rho^{eg}
- \rho^{ge} {\hat V} \right] \nonumber \\
\left(\frac{\partial}{\partial t} -\frac{i\hbar}{M}
\frac{\partial}{\partial q} \frac{\partial}{\partial z}\right)
\rho^{eg}
&+&\left(\frac{\gamma_{eff}}{2}-i\tilde{\delta}(z,q)\right)\rho^{eg}
= -\frac{i}{\hbar} \left[ {\hat V} \rho^{gg}
- \rho^{ee} {\hat V} \right] \nonumber \\
\left(\frac{\partial}{\partial t} -\frac{i\hbar}{M}
\frac{\partial}{\partial q} \frac{\partial}{\partial z}\right)
\rho^{ee}
&+&\left(\frac{\gamma_{eff}}{2}+i\tilde{\delta}(z,-q)\right)\rho^{ge}
= -\frac{i}{\hbar} \left[ {\hat V}^{\dagger} \rho^{ee} - \rho^{gg}
{\hat V}^{\dagger} \right]\,,
\end{eqnarray}
with $z = (z_1+z_2)/2$, $q = z_1-z_2$, and the function
$\tilde{\delta}(z,q)$: $$\tilde{\delta}(z,q) = \delta
-(U_e-U_g)\frac{1+\cos(2kz)\cos(kq)}{2\hbar}
+(U_e+U_g)\frac{\sin(2kz)\sin(kq)}{2\hbar}.$$ The spontaneous income
part to the ground state ${\hat \gamma}$ in coordinate
representation for two-level model has simple form :
\begin{eqnarray}\label{sp_income}
{\hat \gamma}\{\rho^{ee}\} &=& {\tilde{\gamma}}(q) \rho^{ee} \nonumber \\
{\tilde{\gamma}}(q) &=& 3\,\gamma_{eff}\left(
\frac{\sin(k_1q)}{(k_1q)^3}-\frac{\cos(k_1q)}{(k_1q)^2} \right)
\end{eqnarray}
with $k_1 = 2\pi/\lambda$ is wavevector of emitted photon. The
pumping field induces transitions between the ground and excited
states. This part is described in (\ref{basic_eq}) by operator
\begin{equation}\label{Vpump}
{\hat V} = \Omega/2\, \exp(ik_1 z \cos(\theta))
\end{equation}
with $\Omega$ is Rabi frequency of pumping field and $\theta$ is
angle between the axis $z$ and pumping wave propagation direction.
For the case of orthogonal orientation of the pumping wave
propagation to the dipole trap $\theta = \pi/2$ the equation for
density matrix (\ref{basic_eq}) can be solved numerically by the
method suggested in \cite{Prudnikov2007,Prudnikov2011}. It should be
noted the considered method allows to get steady state solution for
density matrix with taking into account quantum recoil effects as
for atoms in the trap as for nontrapped atoms.

\begin{figure}[t]
\begin{center}
\includegraphics[width= 5.0 in]{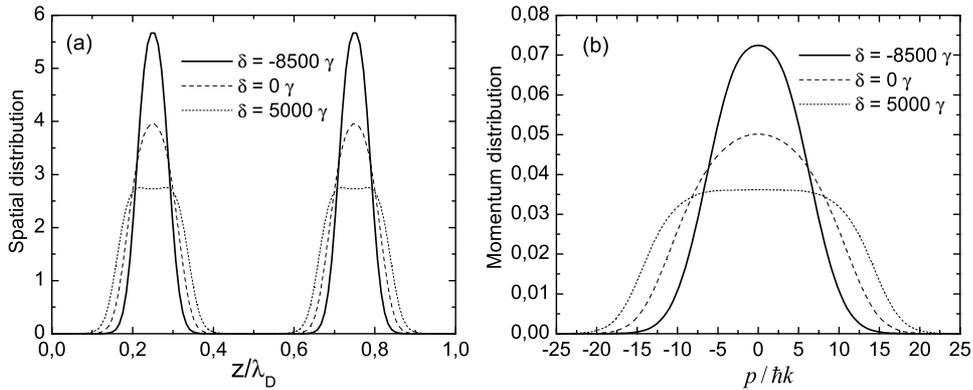}
\end{center}
\caption{\em Spatial (a) and momentum (b) distribution of $^{24}$Mg
atoms in optical dipole trap ($U_g = 200\, \mu K$, $U_e = 292\, \mu
K$) for orthogonal orientation of pumping wave and dipole trap
($\theta = \pi/2$) and pumping field intensity $I \simeq 34$
$W/cm^2$ ($\Omega/\gamma = 20000$) and different detunings.}
\label{mg_dist}
\end{figure}

The figure \ref{mg_dist} shows spatial and momentum distribution of
Mg atoms in the optical dipole trap for orthogonal orientation of
pumping wave ($\theta = \pi/2$), for pumping field
intensity $I \simeq 34$ $W/cm^2$ ($\Omega/\gamma = 20000$) and
different detunings.

\begin{figure}[t]
\begin{center}
\includegraphics[width= 6.0 in]{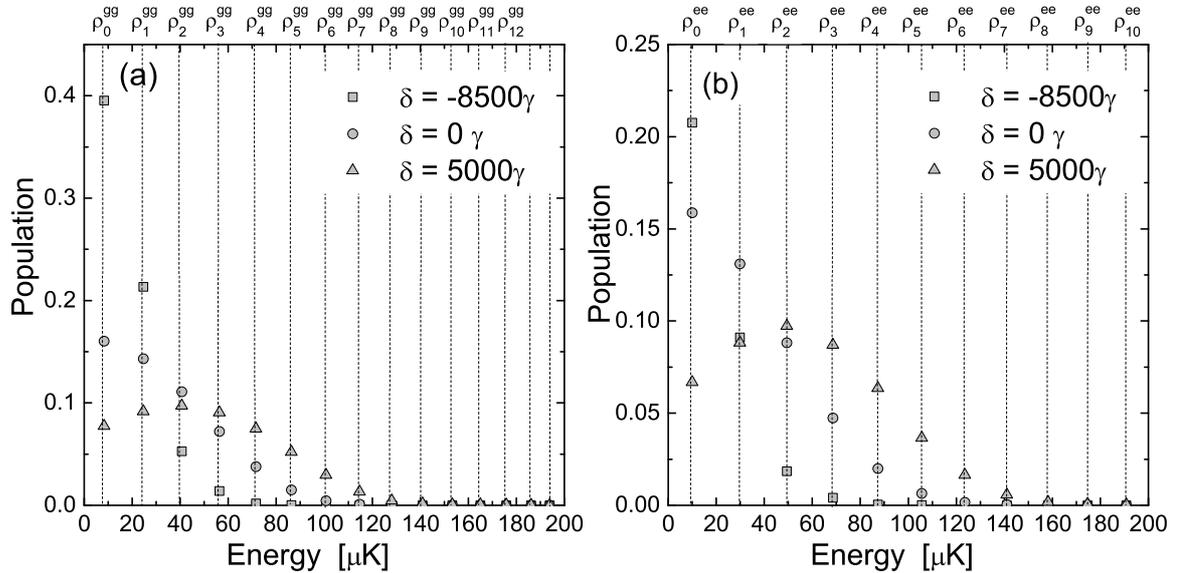}
\end{center}
\caption{\em Vibration energy levels population distribution in the
ground (a) and excited (b) states of $^{24}$Mg atoms in the dipole
trap ($U_g = 200\, \mu K$, $U_e = 292\, \mu K$) for orthogonal
orientation of pumping wave and dipole trap ($\theta = \pi/2$).
Pumping field intensity $I \simeq 34$ $W/cm^2$ ($\Omega/\gamma =
20000$).} \label{mg_popdist}
\end{figure}

The obtained numerical solution for steady state density matrix
${\hat \rho}(z_1,z_2)$ contains whole information on internal and
external states of atoms in the trap. In particular one can extract
the population of vibration levels in the ground and excited states:
\begin{eqnarray}
\rho^{ee}_n &=& \int
\psi^{*(e)}_n(z_1)\rho^{ee}(z_1,z_2)\psi^{(e)}_n(z_2) dz_1 dz_2 \, ,
\nonumber \\
\rho^{gg}_n &=& \int
\psi^{*(g)}_n(z_1)\rho^{gg}(z_1,z_2)\psi^{(g)}_n(z_2) dz_1 dz_2
\end{eqnarray}
where $\psi^{(e,g)}_n(z)$ are n-th vibration level eigenfunctions.
The distribution of vibration levels population in the ground and
excited states for parameters of figure \ref{mg_dist} are shown on
figure \ref{mg_popdist}.

\begin{figure}[t]
\begin{center}
\includegraphics[width= 5.3 in]{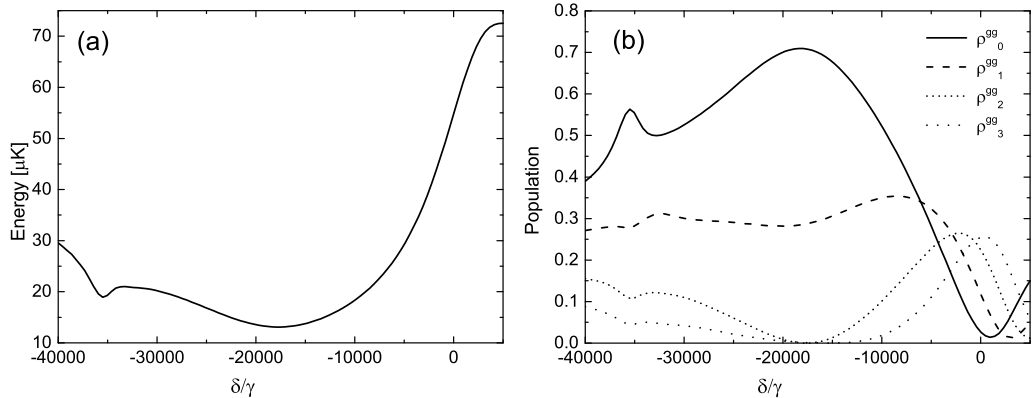}
\end{center}
\caption{\em The energy (a) of cooled $^{24}$Mg atoms in the dipole
trap and population of the lowest vibration levels (b) as function
of pumping field detuning for $I \simeq 0.34$ $W/cm^2$
($\Omega/\gamma = 2000$) obtained by direct numerical solution of
eq.(\ref{basic_eq}). ($\theta = \pi/2$)} \label{model1_dd}
\end{figure}

\begin{figure}[t]
\begin{center}
\includegraphics[width= 3.3 in]{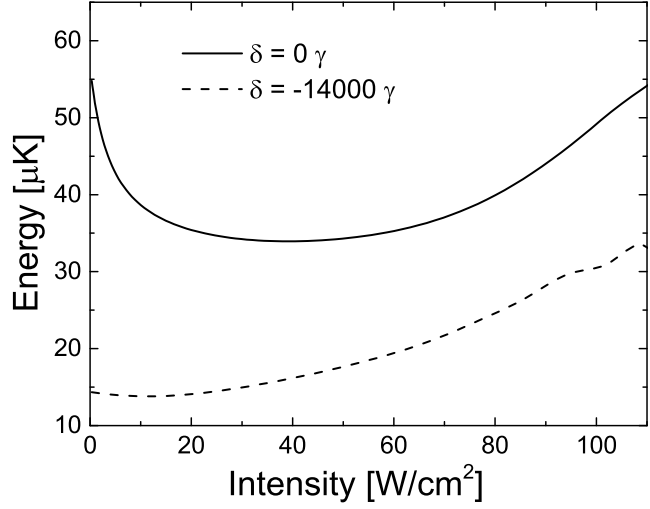}
\end{center}
\caption{\em The energy of cooled $^{24}$Mg atoms in the dipole trap
($\theta = \pi/2$) obtained by direct numerical solution of
eq.(\ref{basic_eq}) as function of pumping field intensity for
different detunings.} \label{model1_Om}
\end{figure}

The energy of cooled atoms can be found by different way. First of
all one can use the relation for the temperature of cooled atoms in
the well known form
\begin{equation}\label{Eclass0}
k_B T = <p^2>/M\,,
\end{equation}
with $<p^2> = Tr\{{\hat p}^2\, {\hat \rho}\}$. This relation
neglects the atom localization effects in the optical potential. The
most accurate relation for energy is expressed by the following
averaging:
\begin{equation}\label{Eclass}
E = Tr\left\{ \left( \frac{{\hat p}^2}{2M}\, + \hbar
\omega_e(z)\right){\hat \rho^{(ee)}}\right\} +Tr\left\{ \left(
\frac{{\hat p}^2}{2M}\, + \hbar \omega_g(z)\right){\hat
\rho^{(gg)}}\right\}\,.
\end{equation}
As an alternative way one can find the average energy over the
vibration states
\begin{equation}\label{E_vibr}
E = \sum_n E^{(e)}_n \, \rho^{(ee)}_n +E^{(g)}_n \,
\rho^{(gg)}_n.
\end{equation}
For considered parameters all above definitions give very close
values that denotes the main contribution to energy are given by
atoms on the lowest vibration energy levels in the region where the
optical potential has close to parabola shape. The energy of Mg
atoms for different detuning is shown on figure \ref{model1_dd}. The
total population of excited vibration level states here do not
exceed $2\%$ and are not shown on figure \ref{model1_dd}(b). In the
region of detuning $\delta > -6000 \gamma$ we find inversion of the
lowest vibration levels population resulting to energy growth. For
the higher intensity of pumping field this effect is also exists and
moves to larger detuning area. The energy of atoms as function of
pumping field intensity for different detunings is shown on figure
(\ref{model1_Om}).

\subsection{Decomposition on vibration states model}
As we see from the simulations above based on numerical solution of
basic equation for atomic density matrix (\ref{basic_eq}) for the
considered parameters Mg atoms can be cooled and well localized in
the dipole trap. Thus for description of laser coolling and laser
cooling time we can also apply an alternative approach based on
decomposition of atom density matrix over vibration level states.
\begin{equation}
{\hat \rho} = \left(
\begin{array}{cc}
\left[%
  \begin{array}{ccc}
    {\rho}^{ee}_{00} & {\rho}^{ee}_{01} & ... \\
    {\rho}^{ee}_{10} & {\rho}^{ee}_{11} & ... \\
    ... & ... & {\rho}^{ee}_{nm} \\
  \end{array}\right] &
  \left[%
  \begin{array}{ccc}
    {\rho}^{eg}_{00} & {\rho}^{eg}_{01} & ... \\
    {\rho}^{eg}_{10} & {\rho}^{eg}_{11} & ... \\
    ... & ... & {\rho}^{eg}_{nm} \\
  \end{array}\right] \\
    \left[%
  \begin{array}{ccc}
    {\rho}^{ge}_{00} & {\rho}^{ge}_{01} & ... \\
    {\rho}^{ge}_{10} & {\rho}^{ge}_{11} & ... \\
    ... & ... & {\rho}^{ge}_{nm} \\
  \end{array}\right] &
    \left[%
  \begin{array}{ccc}
    {\rho}^{gg}_{00} & {\rho}^{gg}_{01} & ... \\
    {\rho}^{gg}_{10} & {\rho}^{gg}_{11} & ... \\
    ... & ... & {\rho}^{gg}_{nm} \\
  \end{array}\right] \\
\end{array}%
\right)
\end{equation}
The equation for these components takes the form:
\begin{equation}\label{vibr_eq}
\frac{\partial }{\partial t}{\hat \rho} =
-\frac{i}{\hbar}\left[{\hat H}_0+{\hat W},{\hat \rho} \right]
+\Gamma\{{\hat \rho}\}
\end{equation}
with Hamiltonian
\begin{equation}
{\hat H}_0 = \left(
\begin{array}{cc}
\left[%
  \begin{array}{ccc}
    E^{(e)}_{0} - \hbar\delta & 0 & ... \\
    0 & E^{(e)}_{1} - \hbar\delta & ... \\
    ... & ... & E^{(e)}_{n} - \hbar\delta \\
  \end{array}\right] &
  \left[%
  \begin{array}{ccc}
    0 & 0 & ... \\
    0 & 0 & ... \\
    ... & ... & 0 \\
  \end{array}\right] \\
    \left[%
  \begin{array}{ccc}
    0 & 0 & ... \\
    0 & 0 & ... \\
    ... & ... & 0 \\
  \end{array}\right] &
    \left[%
  \begin{array}{ccc}
    E^{(g)}_0 & 0 & ... \\
    0 & E^{(g)}_1 & ... \\
    ... & ... & E^{(g)}_n \\
  \end{array}\right] \\
\end{array}%
\right)
\end{equation}
where $E^{(e)}_n$ and $E^{(g)}_n$ are the vibration levels energy of
the excited and the ground states. The pumping light field to atom
interaction part has nondiagonal block elements only:
\begin{equation}
W^{eg}_{nm} = \langle e,n| {\hat V} |g,m \rangle  = \int
\psi^{*(e)}_n(z) {\hat V} \psi^{(g)}_m(z) dz \, .
\end{equation}
with ${\hat V}$ is defined in (\ref{Vpump}). The spontaneous relaxation
part $\Gamma\{{\hat \rho}\}$ has a standard form
\begin{equation}
\Gamma\{ {\hat \rho} \} = -\frac{\gamma_{eff}}{2} \left\{ |e,n
\rangle \langle e,n|,{\hat \rho} \right\} + {\hat \gamma} \left\{
{\hat \rho} \right\}
\end{equation}
with ${\hat \gamma} \left\{ {\hat \rho} \right\}$ describes income
to the ground vibration states has the following matrix elements:
\begin{eqnarray}
{\hat \gamma} \left\{ {\hat \rho} \right\}_{nm} &=& \sum_{\nu \mu}
\Gamma^{\nu \mu}_{nm} {\rho}^{ee}_{\nu \mu} \nonumber \\
\Gamma^{\nu \mu}_{nm} & = & \int
\psi^{*(g)}_n(z_1)\psi^{(e)}_{\nu}(z_1) {\tilde{\gamma}}(z_1-z_2)
\psi^{*(e)}_n(z_2)\psi^{(g)}_{\nu}(z_2) dz_1 dz_2
\end{eqnarray}
and ${\tilde{\gamma}}(q)$ is defined in (\ref{sp_income}).  Thus in
the equation (\ref{vibr_eq}) we take into account the evolution of
diagonal elements of density matrix (vibration levels population)
and nondiagonal elements as well. However, compare to exact
numerical solution described above we should restrict our
consideration by limited number of vibration levels neglecting
tunneling effects and above barrier motion. Here bellow we consider
10 vibration levels on the ground and excited states.

\begin{figure}[t]
\begin{center}
\includegraphics[width= 6.0 in]{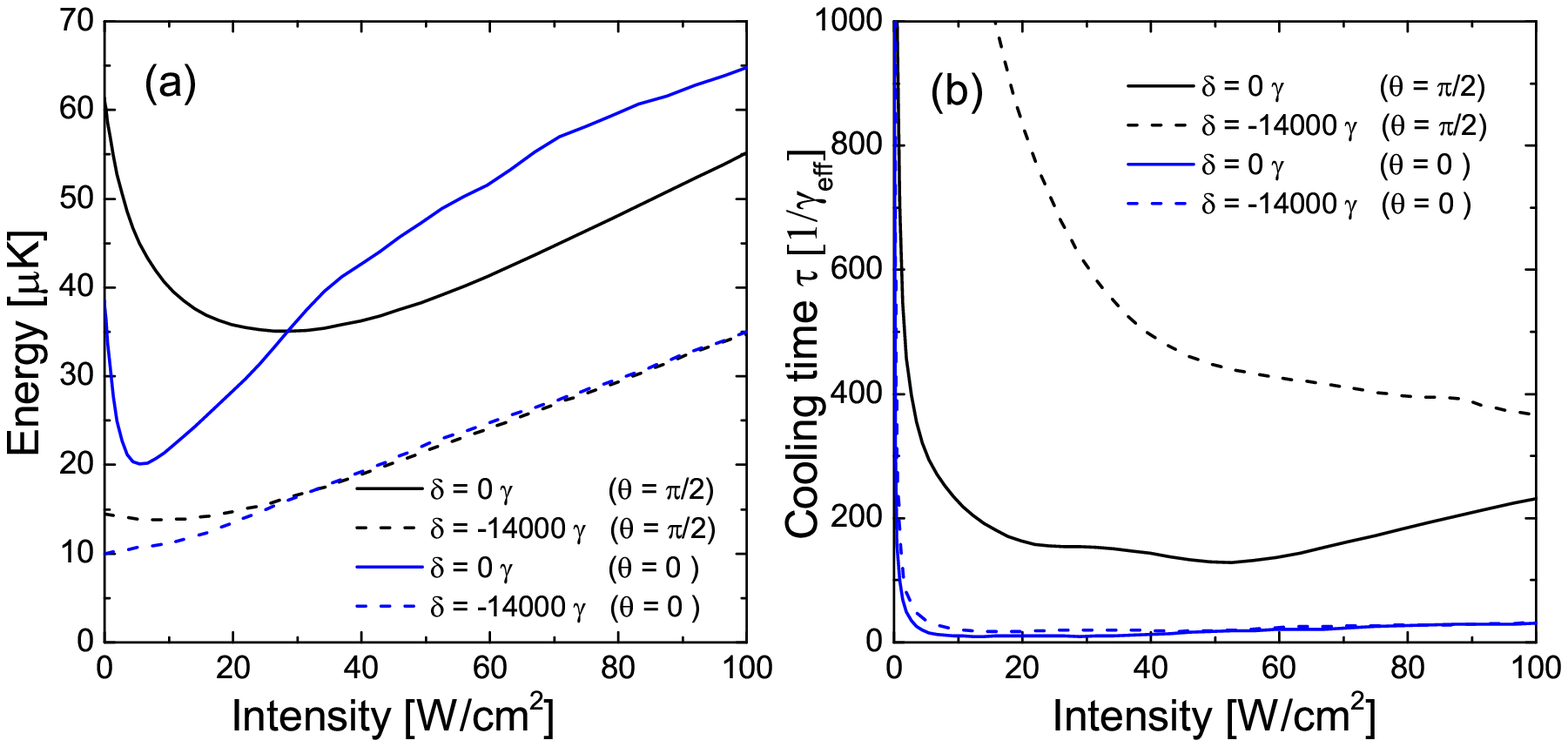}
\end{center}
\caption{\em Energy of cooled $^{24}$Mg atoms in the dipole trap (a)
and cooling time $\tilde{\tau}$ (b) in considered optical dipole
trap for different orientation of pumping wave and dipole trap and
different detunings obtained by solution based on decomposition on
vibration state model as function of pumping field intensity.}
\label{model2_Om}
\end{figure}

The atom steady state energy as function of pumping field intensity
is shown on figure \ref{model2_Om}(a). Even the model is simplified
it gives the cooling energy result close to the direct numerical
solution of eq.(\ref{basic_eq}) (see figure \ref{model1_Om}) at
$\theta =\pi/2$. The difference appears at large pumping field
intensity where the populations of the top vibration levels are not
negligible and tunneling effects can not be neglected, i.e. far from
the field parameters required for cooling to minimum energy.
Additionally this model allows to solve dynamical problem and
estimate the cooling time. To find the cooling time we assume the
atoms populate the highest vibrational energy level of the ground
state optical potential at $t = 0$. The time evolution of vibration
levels population has a complex dependence. We fit
${\rho}^{gg}_{00}(t)$ by exponential function of the form
${\rho}^{gg}_{00}(t) = a - b \, \mbox{exp}(-t/{\tau})$ with $\tau$
describes the cooling time. Additionally we note, the energy of
cooled atoms does not depend on parameter $\gamma_{eff}$ (i.e. on
quench field intensity) in the range of our simulations $\gamma <
\gamma_{ef} < 100 \gamma$, while the cooling time $\tau$ is
inversely proportional to $\gamma_{eff}$ in considered model. This
allows us to represent the cooling time $\tau$ in the more general
form through dimensionless value $\tilde{\tau}$ figure
\ref{model2_Om}(b)
\begin{equation}
\tau = \tilde{\tau} /\gamma_{eff}.
\end{equation}

\begin{figure}[t]
\begin{center}
\includegraphics[width= 6.8 in]{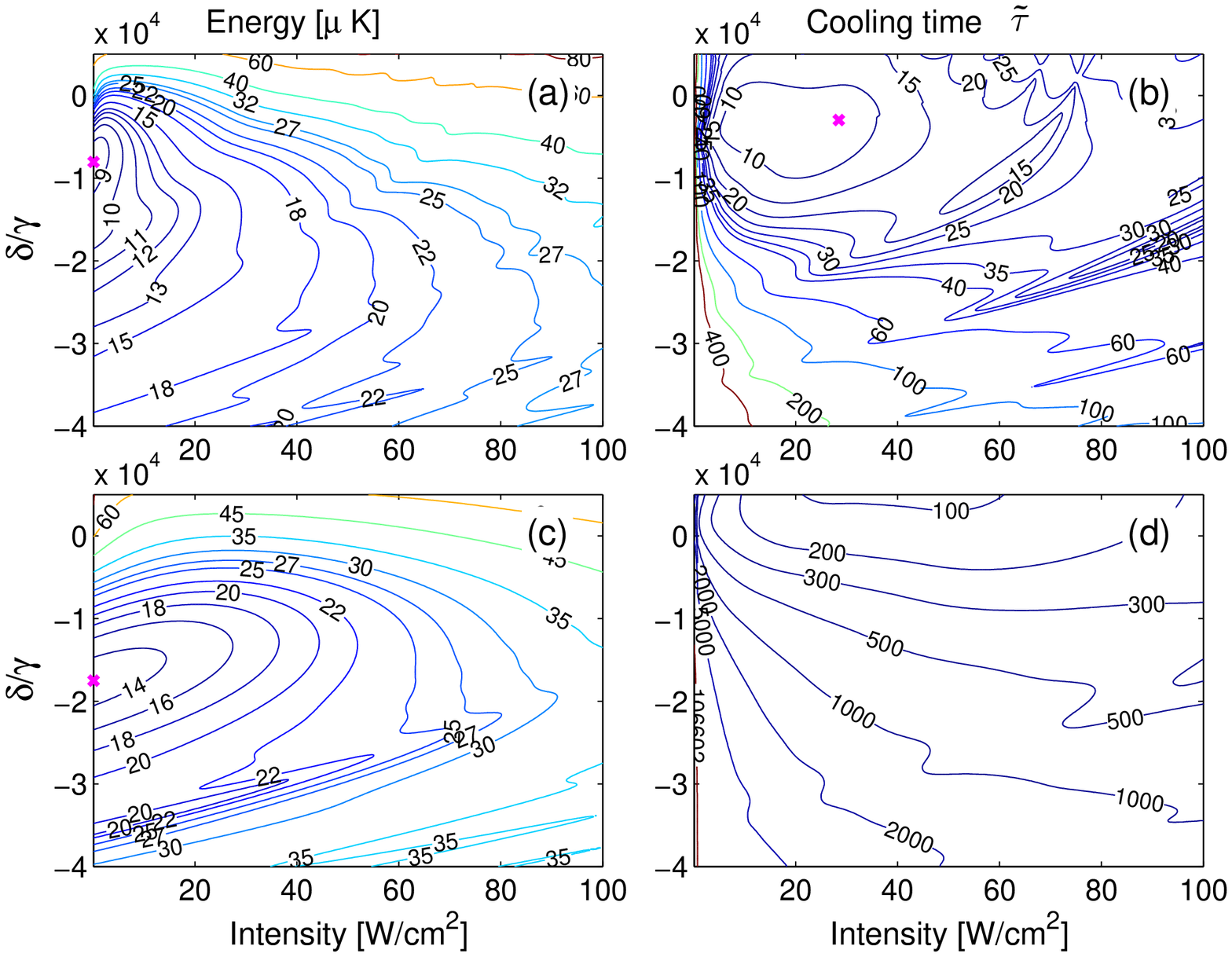}
\end{center}
\caption{\em The energy of cooled $^{24}$Mg atoms ((a) for $\theta =
0$ and (c) for $\theta = \pi/2$) in $\mu K$ and the cooling time
$\tilde{\tau}$ in dimensionless units ((b) for $\theta = 0$ and (d)
for $\theta = \pi/2$) in the dipole trap as function of pumping wave
intensity and detunings obtained by the model based on decomposition
of atomic density matrix on vibration state. Magenta cross define
optimal parameters for minimum laser cooling energy of atoms and
minimum cooling time for considered range of parameters.}
\label{results_2dOm}
\end{figure}

The cooling time and the energy of atoms as function of two
parameters: pumping field detuning and intensity are shown on figure
\ref{results_2dOm}. The conditions for minimum cooling energy do not
coincide with conditions for fast cooling. For orientation angle
$\theta = 0$ the minimum of cooling energy $E_{min} = 8.4 \mu K$ is
reached for pumping field intensity $I \simeq 0.1 W/cm^2$ and
detuning $\delta/\gamma \simeq -8050$, while the minimum cooling
time $\tau\gamma_{eff} \simeq 7.4$ is reached at $I \simeq 28.5
W/cm^2$ and detuning $\delta/\gamma \simeq -2987$ (figure
\ref{results_2dOm}(a,b)).

For orthogonal orientation $\theta = \pi/2$ the minimum energy
$E_{min} = 13.2 \mu K$ is reached for  $I \simeq 0.1 W/cm^2$ and
detuning $\delta/\gamma \simeq -17500$, but the cooling time for
these parameters is extremely large $\tau\gamma_{eff} \simeq 2
\cdot 10^5$. For field parameters providing a reasonable cooling
time $\tau\gamma_{eff} <100$ the energy above $60 \mu K$ can be
reached only (figure \ref{results_2dOm}(c,d)).

\section{Conclusion}
We study laser cooling of $^{24}Mg$ atoms in the dipole trap with
pumping field resonant to narrow $(3s3s)\,^1S_0 \rightarrow \,
(3s3p)\,^{3}P_1$ ($\lambda = 457$ nm) optical transition and quench
field resonant to $(3s3p)\,^3P_1 \rightarrow \, (3s4s)\,^{1}S_0$.
The effect of quenching we consider as widening of optical
transition to $\gamma_{eff}$ only. We find the semiclassical model
can not be used for description of laser cooling in this scheme. We
suggest quantum models. The first one is based on the direct
numerical solution of quantum kinetic equation for atom density
matrix and the second is simplified model is based on decomposition
of atom density matrix over vibration states in the dipole trap. The
second model has limitations and describes the cooling of atoms on
the lowest vibration levels only. The results of this model is
differ from exact numerical solution for enough high intensity of
pumping field (above 50$W/cm^2$ or Rabi above $25000 \gamma$) when
populations of the top vibration levels are not negligible, i.e.
tunneling effects and above barrier motion of atoms can not be
neglected. Nevertheless the simplified model well describes the
laser cooling of atoms cooled to minimum energy. Additionally, the
simplified model allows to estimate the cooling time. The parameters
of pumping field for cooling to minimum energy do not coincide with
conditions for fast cooling. We find parameters that allow cooling
the atoms at reasonable cooling time $\tau \simeq 10 / \gamma_{eff}$
(i.e. $\tau \simeq 0.5\, ms$ at $\gamma_{eff} = 100\gamma$) to
energy $E\simeq 12 \, \mu K$.

 In the considered models the steady state solution
do not depends on $\gamma_{eff}$ while the cooling time is inversely
proportional to $\gamma_{eff}$. The difference may appears for more
complex model that takes into account pumping to level
$(3s4s)\,^{1}S_0$ by quench field. We consider this question in the
next paper.

The work was supported by Russian Science Foundation (project N
16-12-00054). V.I.Yudin acknowledges the support of the Ministry of
Education and Science (3.1326.2017).


\begin{thebibliography}{99}
\bibitem{Dalibard1989} J. Dalibard and C. Cohen-Tannoudji, J. Opt. Soc. Am. B {\bf
6,} 2023-2045 (1989).
\bibitem{Aspect1994}J. Lawall, F. Bardou, B. Saubamea, K. Shimizu, M. Leduc, A.
Aspect, and C. Cohen-Tannoudji, ��Two-dimensional subrecoil laser
cooling,�� Phys. Rev. Lett. {\bf 73,} 1915 (1994).
\bibitem{Adams1995} C. S. Adams, H. J. Lee, N. Davidson, M. Kasevich,
and S. Chu, ��Evaporative cooling in a crossed dipole trap,�� Phys.
Rev. Lett. {\bf 74,} 3577�3580 (1995).
\bibitem{Kasevich1992} M. Kasevich and S. Chu, ��Laser cooling below a photon recoil
with three-level atoms,�� Phys. Rev. Lett. {\bf 69,} 1741 (1992).
\bibitem{ctannoudji1995} J. Reichel, F. Bardou, M. Ben Dahan, E. Peik, S. Rand, C.
Salomon, and C. Cohen-Tannoudji, ��Raman cooling of cesium below 3
nK: new approach inspired by Le�vy flight statistics,�� Phys. Rev.
Lett. {\bf 75,} 4575 (1995).
\bibitem{Rasel2015} A.P. Kulosa, D. Fim, K.H. Zipfel, S. Ruhmann, S.
Sauer, N. Jha, K. Gibble, W. Ertmer, E.M. Rasel, M.S. Safronova,
U.I. Safronova, S.G. Porsev, arXiv: 1508.01118v1, physics.atom-ph, 5
Aug 2015.
\bibitem{Rasel2012} M. Riedmann, H. Kelkar, T. Wubbena, A. Pape, A.
Kulosa, K. Zipfel, D. Fim, S. Ruhmann, J. Friebe, W. Ertmer, and E.
Rasel, Phys. Rev. A {\bf 86,} 043416 (2012).
\bibitem{Prudnikov2016} O.N. Prudnikov, D. V. Brazhnikov, A. V. Taichenachev, V. I. Yudin, V I Yudin, A N Goncharov,
Quantum Electronics, v. 46, Issue 7, p. 661-667, (2016)
\bibitem{holberg2001} E.A. Curtis, C.W. Oates, and L.Holberg Phys.
Rev. A {\bf 64,} 031403 (2001)
\bibitem{Rasel2001} T. Binnewies, G. Wilpers, U. Sterr, J. Helmcke,
T.E. Mehlst\"{a}bler, E.M. Rasel, and W. Ertmer Phys. Rev. Lett.
{\bf 87,} 123002 (2001)
\bibitem{Rasel2003} T.E. Mehlst\"{a}bler, J. Keupp, A.Douillet,
N. Rehbein, E.M. Rasel J.Opt.B: Quantum Semiclass. Opt. {\bf 5,}
S183 (2003)
\bibitem{Ivanov2011} V.V. Ivanov, S. Gupta Phys.Rev. A {\bf 84,
} 063417 (2011)
\bibitem{Juha1990} J. Javanainen Phys. Rev. A {\bf 44,} 5857-5880 (1990)
\bibitem{Prudnikov2004} O.N. Prudnikov, A. V. Taichenachev, A. M. Tumaikin and V. I. Yudin, JETP {\bf 98,} 438-454 (2004)
\bibitem{Castin1991} Y. Castin and J. Dalibard Europhys. Lett., {\bf 14,} 761-766 (1991)
\bibitem{Prudnikov2007} O.N. Prudnikov, A.V. Taichenachev, A.V. Tumaikin, V.I. Yudin, Phys. Rev. A 75, 023413 (2007)
\bibitem{Prudnikov2011} O. N. Prudnikov, R.Ya. Ilenkov, A. V. Taichenachev, A. M. Tumaikin, and V. I. Yudin, JETP v.112, pp.939-945 (2011)


\end{thebibliography}
\end{document}